\documentclass[
    aip,
    rsi,
    reprint,
    noeprint
]{revtex4-1}
\usepackage{amsmath,amsfonts,amscd,amssymb}
\usepackage{cancel}
\DeclareMathOperator*{\argmax}{arg\rm{}max}

\usepackage{graphicx}        
\graphicspath{{./figures/}}
\usepackage{latexsym}
\usepackage{overpic}
\usepackage{epstopdf}
\usepackage{rotating,color}

\usepackage{lmodern,textcomp}
\usepackage{listings}
\usepackage{url}
\usepackage{paralist}
\usepackage{bm}
\usepackage{booktabs}
\usepackage{hyperref}

\newcommand{\norm}[1]{\left\lVert#1\right\rVert}

\newcommand{\bx}{\mathbf{x}}

\begin{document}

\begin{abstract}
Big data has become a critically enabling component of emerging mathematical methods aimed at the automated discovery of dynamical systems, where first principles modeling may be intractable.
%
%
However, in many engineering systems, abrupt changes must be rapidly characterized based on limited, incomplete, and noisy data.
Many leading automated learning techniques rely on unrealistically large data sets and it is unclear how to leverage prior knowledge effectively to re-identify a model after an abrupt change.
In this work, we propose a conceptual framework to recover parsimonious models of a system in response to abrupt changes in the low-data limit.
First, the abrupt change is detected by comparing the estimated Lyapunov time of the data with the model prediction.
Next, we apply the sparse identification of nonlinear dynamics (SINDy) regression to update a previously identified model with the fewest changes, either by addition, deletion, or modification of existing model terms.
We demonstrate this sparse model recovery on several examples for abrupt system change detection in periodic and chaotic dynamical systems.
Our examples show that sparse updates to a previously identified model perform better with less data, have lower runtime complexity, and are less sensitive to noise than identifying an entirely new model.
The proposed abrupt-SINDy architecture provides a new paradigm for the rapid and efficient recovery of a system model after abrupt changes.

\end{abstract}

\author{Markus Quade}
\email{mquade@uni-potsdam.de}
\homepage[\\All code for this work is publicly available on Github: ]{https://github.com/Ohjeah/sparsereg}
\author{Markus Abel}
\affiliation{Universit\"at Potsdam, Institut f\"ur Physik und Astronomie, Karl-Liebknecht-Stra{\ss}e 24/25, 14476 Potsdam, Germany}
\affiliation{Ambrosys GmbH, David-Gilly Stra{\ss}e 1, 14469 Potsdam, Germany}

\author{J. Nathan Kutz}
\affiliation{Department of Applied Mathematics, University of Washington, Seattle, WA 98195, USA}

\author{Steven L. Brunton}
\affiliation{Department of Mechanical Engineering, University of Washington, Seattle, WA 98195, USA}

\title{Sparse Identification of Nonlinear Dynamics for Rapid Model Recovery}

\keywords{Dynamical systems, Chaos, Data-driven models, Machine learning, Sparse optimization}
\pacs{02.60.-x, 02.60.Ed, 05.45.-a, 05.45.Pq, 05.45.Tp, 07.05.Kf}

\maketitle

\noindent \textbf{Dynamical systems modeling is a cornerstone of modern mathematical physics and engineering.  The dynamics of many complex systems (e.g., neuroscience, climate, epidemiology, etc.) may not have first-principles derivations, and researchers are increasingly using data-driven methods for system identification and the discovery of dynamics.  Related to  discovery of dynamical systems models from data is the \emph{recovery} of these models following abrupt changes to the system dynamics.  In many domains, such as aviation, model recovery is mission critical, and must be achieved rapidly and with limited noisy data.  This paper leverages recent advances in sparse optimization to identify the fewest terms required to recover a model, introducing the concept of \emph{parsimony of change}.  In other words, many abrupt system changes, even catastrophic bifurcations, may be characterized with relatively few changes to the terms in the underlying model.  In this work, we show that sparse optimization enables rapid model recovery that is faster, requires less data, is more accurate, and has higher noise robustness than the alternative approach of re-characterizing a model from scratch.  
}


\section{Introduction}
The data-driven discovery of physical laws and dynamical systems is poised to revolutionize how we model, predict, and control physical systems.
Advances are driven by the confluence of big data, machine learning, and modern perspectives on dynamics and control.
However, many modern techniques in machine learning (e.g., neural networks) often rely on access to massive data sets, have limited ability to generalize beyond the attractor where data is collected, and do not readily incorporate known physical constraints.
These various limitations are framing many state-of-the-art research efforts around learning algorithms~\cite{goodfellow2016deep}, especially as it pertains to generalizability, limited data and {\em one-shot learning}~\cite{fei2006one,vinyals2016matching,delahunt2018putting}.
Such limitations also frame the primary challenges and limitations associated with data-driven discovery for real-time control of strongly nonlinear, high-dimensional, multi-scale systems with abrupt changes in the dynamics.
Whereas traditional methods often require unrealistic amounts of training data to produce a viable model, this work focuses on methods that take advantage of prior experience and knowledge of the physics to dramatically reduce the data and time required to characterize dynamics.
Our methodology is similar in philosophy to the machine learning technique of {\em transfer learning}~\cite{pan2010survey}, which allows networks trained on one task to be efficiently adapted to another task.
Our architecture is designed around the goal of rapidly extracting parsimonious, nonlinear dynamical models that identify only the fewest important interaction terms so as to avoid overfitting.

There are many important open challenges associated with data-driven discovery of dynamical systems for real-time tracking and control.
When abrupt changes occur in the system dynamics, an effective controller must rapidly characterize and compensate for the new dynamics, leaving little time for recovery based on limited data~\cite{Brunton2015amr}.
The primary challenge in real-time model discovery is the reliance on large quantities of training data.
A secondary challenge is the ability of models to generalize beyond the training data, which is related to the ability to incorporate new information and quickly modify the model.
Machine learning algorithms often suffer from overfitting and a lack of interpretability, although the application of these algorithms to physical systems offers a unique opportunity to enforce known symmetries and physical constraints (e.g. conservation of mass).
Inspired by biological systems, which are capable of extremely fast adaptation and learning based on very few trials of new information~\cite{rankin2004invertebrate,whitlock2006learning,johansen2011molecular}, we propose model discovery techniques that leverage an \emph{experiential framework}, where known physics, symmetries, and conservation laws are used to rapidly infer model changes with limited data.

\vspace{-.1in}
\subsection{Previous work in system identification}
\vspace{-.1in}
There are a wealth of regression techniques for the characterization of system dynamics from data, with varying degrees of generality, accuracy, data requirements, and computational complexity.
Classical linear model identification algorithms include Kalman filters~\cite{Kalman1960jfe,Kalman1965AAC,gershenfeld1999nature}, the eigensystem realization algorithm (ERA)~\cite{Juang1985jgcd}, dynamic mode decomposition (DMD)~\cite{Schmid2010jfm,Rowley2009jfm,Tu2014jcd,Kutz2016book}, and autoregressive moving average (ARMA) models~\cite{Akaike1969annals,Brockwell2017}, to name only a few.
The resulting linear models are ideal for control design, but are unable to capture the underlying nonlinear dynamics or structural changes.
Increasingly, machine learning is being used for nonlinear model discovery.
Neural networks have been used for decades to identify nonlinear systems~\cite{gonzalez1998identification}, and are experiencing renewed interest because of the ability to train deeper networks with more data~\cite{goodfellow2016deep,Yeung2017arxiv,Takeishi2017nips,Wehmeyer2017arxiv,Mardt2017arxiv,Lusch2017arxiv} and the promise of transformations that linearize dynamics via the Koopman operator~\cite{Koopman1931pnas,Mezic2005nd}.
Neural networks show good capacity to recover the dynamics in a so-called ``model-free'' way~\cite{Lukosevicius2009,Lu2017}. These methods are also known as ``reservoir computers'', ``liquid state machines'', or ``echo state networks'', depending on the context.
However, a real-time application is unrealistic, and the output is generally not analytically interpretable.
In another significant vein of research, genetic programming~\cite{dantzig1985mathematical,koza1992genetic} is a powerful bio-inspired method that has successfully been applied to system identification \cite{Bongard2007pnas,Schmidt2009science,schmidt2011automated,LaCava2016b}, time-series prediction \cite{LaCava2016a,Quade2016} and control \cite{Gout2018,Duriez2017}. However, evolutionary methods in their pure form, including genetic programming, are computationally complex and thus are not suitable for real-time tracking.

Recently, \emph{interpretability} and \emph{parsimony} have become important themes in nonlinear system identification~\cite{Bongard2007pnas,Schmidt2009science}.
A common goal now is to identify the fewest terms required to represent the nonlinear structure of a dynamical system model while avoiding overfitting~\cite{Brunton2016pnas}.
Symbolic regression methods~\cite{voss1998,Schmidt2009science,McConaghy2011,Brunton2016pnas} are generally appealing for system identification of structural changes, although they may need to be adapted to the low-data limit and for faster processing time.
Nonparametric additive regression models~\cite{abel2005additive,voss1999,abel2004} require a backfitting loop which allows general transformations, but may be prohibitively slow for real-time applications.
Generalized linear regression methods are slightly less general but can be brought to a fast evaluation and sparse representation \cite{McConaghy2011,Brunton2016pnas}.
These leading approaches to identify dynamical equations from data usually rely on past data and aim at reliable reproduction of a stationary system, i.e. when the underlying equations do not change in the course of time~\cite{Schmidt2009science,
abel2004,Brunton2016pnas}.

\vspace{-.1in}
\subsection{Contributions of this work}
\vspace{-.1in}
In this work, we develop an adaptive modification of the sparse identification of nonlinear dynamics (SINDy) algorithm~\cite{Brunton2016pnas} for real-time recovery of a model following abrupt changes to the system dynamics.   
We refer to this modeling framework as \emph{abrupt-SINDy}.
Although this is not the only approach for real-time change detection and recovery, parsimony and sparsity are natural concepts to track abrupt changes, focusing on the fewest modifications to an existing model.
SINDy already requires relatively small amounts of data~\cite{Kaiser2017arxivB}, is based on fast regression techniques, and has been extended to identify PDEs~\cite{Rudy2017sciadv,Schaeffer2017prsa}, to include known constraints and symmetries~\cite{Loiseau2016arxiv}, to work with limited measurements~\cite{Brunton2017natcomm} and highly corrupted and noisy data~\cite{Tran2016arxiv,Schaeffer2017pre}, to include control inputs~\cite{Brunton2016nolcos,Kaiser2017arxivB},
and to incorporate information criteria to assess the model quality~\cite{Mangan2017prsa}, which will be useful in abrupt model recovery.

\begin{figure*}
\begin{center}
\includegraphics[width=\textwidth]{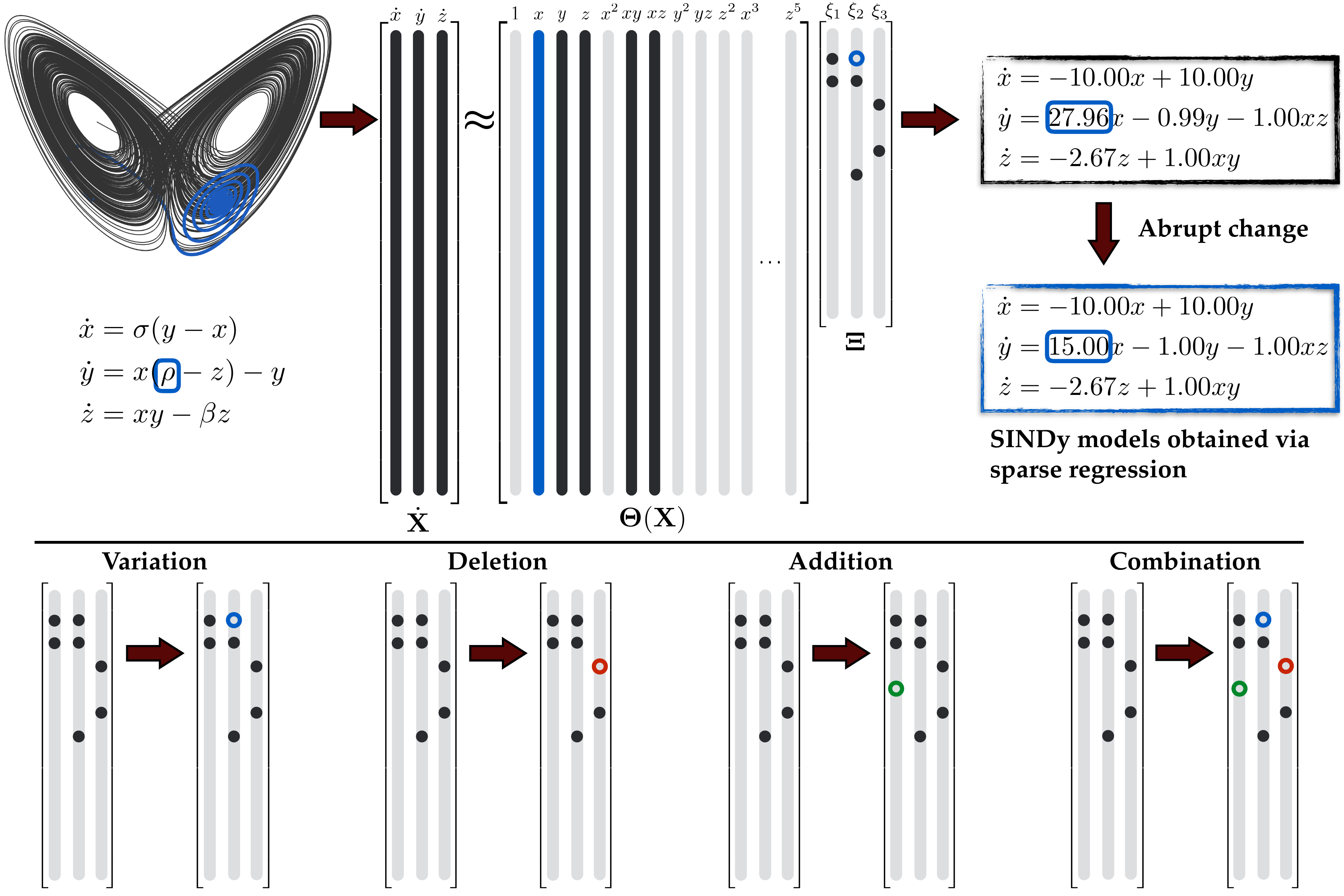}
\vspace{-.25in}
\caption{Schematic overview of abrupt-SINDy method.  (top) Illustration of single variation in a term, giving rise to an abrupt change in the dynamics, from black to blue.  (bottom) Canonical sparse changes, including parameter variation, addition, deletion, or a combination. The data and results for the top panel are from the example in Sec.~\ref{Sec:Lorenz} below.}\label{Fig:Overview}
\vspace{-.2in}
\end{center}
\end{figure*}

Here, we demonstrate that the abrupt-SINDy architecture is capable of rapidly identifying sparse changes to an existing model to recover the new dynamics following an abrupt change to the system.
The first step in the adaptive identification process is to detect a system change using divergence of the prediction from measurements.  Next, an existing model is updated with sparse corrections, including parameter variations, deletions, and additions of terms.  We show that identifying sparse model changes from an existing model requires less data, less computation, and is more robust to noise than identifying a new model from scratch. 
Further, we attempt to maintain a critical attitude and caveat limitations of the proposed approach, highlighting when it can break down and suggesting further investigation.  
The overarching framework is illustrated in Fig.~\ref{Fig:Overview}.

\section{State of the art}

Recently, sparse regression in a library of candidate nonlinear functions has been used for sparse identification of nonlinear dynamics (SINDy) to efficiently identify a sparse model structure from data~\cite{Brunton2016pnas}.
The SINDy architecture bypasses an intractable brute-force search through all possible models, leveraging the fact that many dynamical systems  of the form
\begin{equation}
\frac{d}{dt}{\bf {x}} = {\bf f}({\bf x})
\end{equation}
have dynamics ${\bf f}$ that are sparse in the state variable ${\bf x}\in\mathbb{R}^n$.
Such models may be identified using a sparsity-promoting regression~\cite{Tibshirani1996lasso,Hastie2009book,James2013book} that penalizes the number of nonzero terms $\xi_{ij}$ in a generalized linear model:
\begin{equation}
    \hat{f_i}({\bf x}) = \sum_{j=1}^p\xi_{ij} \theta_{j}({\bf x}),
    \label{eq:glm}
\end{equation}
where $\theta_j({\bf x})$ form a set of nonlinear candidate functions.
The candidate functions may be chosen to be polynomials, trigonometric functions, or a more general set of functions~\cite{Brunton2016pnas,McConaghy2011}.
With poor choice of the candidate functions $\theta_j$, i.e. if library functions are non-orthogonal and/or overdetermined, the SINDy approach may fail to identify the correct model.

Sparse models may be identified from time-series data, which are collected and formed into the data matrix
\begin{equation}
{\bf X} = \begin{bmatrix} {\bf x}_1 & {\bf x}_2 & \cdots {\bf x}_m\end{bmatrix}^T.
\end{equation}
We estimate the time derivatives using a simple forward Euler finite-difference scheme, i.e. the difference of two consecutive data, divided by the time difference:
\begin{equation}
{{\bf \dot{X}}} = \begin{bmatrix} {{\bf \dot{x}}}_1 & {{\bf \dot{x}}}_2 & \cdots {{\bf \dot{x}}}_m\end{bmatrix}^T.
\end{equation}
This estimation procedure is numerically ill-conditioned if data are noisy, although there are many methods to handle noise which work very well if used correctly \cite{Ahnert-Abel-2007,Chartrand2011isrnam}.  
Noise-robust derivatives were investigated in the original SINDy algorithm~\cite{Brunton2016pnas}.  
Next, we consider a library of candidate nonlinear functions $\boldsymbol{\Theta}({\bf X})$, of the form
\begin{equation}
\boldsymbol{\Theta}({\bf X}) = \begin{bmatrix} \mathbf{1} & {\bf X} & {\bf X}^2 & \cdots & {\bf X}^d  & \cdots &   \sin({\bf X}) & \cdots  \end{bmatrix}.
\end{equation}
%
%
%
Here, the matrix ${\bf X}^d$ denotes a matrix with column vectors given by all possible time-series of $d$-th degree polynomials in the state ${\bf x}$.
The terms in $\boldsymbol{\Theta}$ can be functional forms motivated by knowledge of the physics.  Within the proposed work, they may parameterize a piecewise-affine dynamical model.
Following best practices of statistical learning \cite{Hastie2009book}, to preprocess, we mean-subtract and normalize each column of $\boldsymbol{\Theta}$ to have unit variance.
The dynamical system can now be represented in terms of the data matrices as
\begin{equation}
{{\bf \dot{X}}} \approx \boldsymbol{\Theta}({\bf X})\boldsymbol{\Xi}.
\end{equation}
The coefficients in the column $\boldsymbol{\Xi}_k$ of $\boldsymbol{\Xi}$ determine the active  terms in the $k$-th row of Eq.~\eqref{eq:glm}. %
A parsimonious model has the fewest terms in $\boldsymbol{\Xi}$ required to explain the data.
One option to obtain a sparse model is via convex $\ell_1$-regularized regression:
\begin{equation}
\boldsymbol{\Xi} = \text{argmin}_{\boldsymbol{\Xi}'}\|{\mathbf{\dot{X}}} - \boldsymbol{\Theta}(\mathbf{X})\boldsymbol{\Xi}'\|_2+\gamma \|\boldsymbol{\Xi}'\|_1.
\end{equation}
The hyper parameter $\gamma$ balances complexity and sparsity of the solution.
Sparse regression, such as  LASSO \cite{Tibshirani1996lasso} and sequential thresholded least-squares~\cite{Brunton2016pnas}, improves the robustness of identification for noisy overdetermined data, in contrast to earlier methods~\cite{Wang2011prl} using compressed sensing~\cite{Donoho2006ieeetit,Candes2006picm}.
Other regularization schemes may be used to improve performance, such as the elastic net regression \cite{Li2016}.

In this paper we use the sequentially thresholded ridge regression~\cite{Rudy2017sciadv}, which iteratively solves the ridge regression
\begin{equation}
\boldsymbol{\Xi} = \text{argmin}_{\boldsymbol{\Xi}'}\|{\mathbf{\dot{X}}} - \boldsymbol{\Theta}(\mathbf{X})\boldsymbol{\Xi}'\|_2+\alpha \|\boldsymbol{\Xi}'\|_2.
\end{equation}
and then thresholds any coefficient that is smaller than $\gamma$.
The procedure is repeated on the non-zero entries of $\boldsymbol{\Xi}$ until the model converges. The convergence of the SINDy architecture has been discussed in \cite{Zhang2018}. 
After a sparse model structure has been identified in normalized coordinates, it is necessary to regress onto this sparse structure in the original unnormalized coordinates.
Otherwise, non-physical constant terms appear when transforming back from normalized coordinates due to the mean-subtraction.


In La Cava et al.~\cite{LaCava2016b} the authors pursue a complementary although more computationally intensive idea of adaptive modeling in the context of generalized linear models. Starting from an initial guess for the model, a brute force search is conducted to scan a larger set of candidate functions $\theta \rightarrow \theta \theta'^{\gamma}$, where  $\theta'$ are multiplicative extensions to the initial set of candidate functions and $\gamma$ are real valued exponents. The intended use of this method is the refinement of first-principle based models by discovery of coupling terms. It is possible to combine this refinement with our proposed scheme for dealing with abrupt changes.
In addition, sparse sensors~\cite{Manohar2017csm} and randomized algorithms~\cite{Erichson2016arxivA} may improve speed.


\section{Methods}
The viewpoint of sparsity extends beyond model discovery, and we propose to extend SINDy to identify systems undergoing abrupt changes. 
It may be the case that abrupt model changes will only involve the addition, deletion, or modification of a few terms in the model.
This is a statement of the \emph{parsimony of change}, and indicates that we can use sparse regression to efficiently identify the new or missing terms with considerably less data than required to identify a new model from scratch.
In general, each additional term that must be identified requires additional training data to distinguish between joint effects.
Thus, having only a few changes reduces the amount of data required, making the model recovery more rapid.
This section will describe a procedure that extends SINDy to handle three basic types of model changes:
\newline\indent  {\bf {\it i})   Variation of a term.}  If the structure of the model is unchanged and only the parameters vary, we will perform least-squares regression on the known structure to identify the new parameters.  This is computationally fast, and it is easy to check if the model explains the new dynamics, or if it is necessary to explore possible additions or deletions of terms.
\newline\indent  {\bf {\it ii})   Deletion of a term.}  If the model changes by the removal of a few terms, then SINDy regression can be applied on the sparse coefficients in order to identify which terms have dropped out. %
\newline\indent  {\bf {\it iii})   Addition of a term.}  If a term is added, then SINDy regression will find the sparsest combination of inactive terms that explain the model error.  Since least squares regression scales asymptotically $\mathcal{O}(p^3)$, with $p$ the number of columns in the library, this is computationally less expensive than regression in the entire library.

\begin{figure}[t]
    \includegraphics[width=\columnwidth]{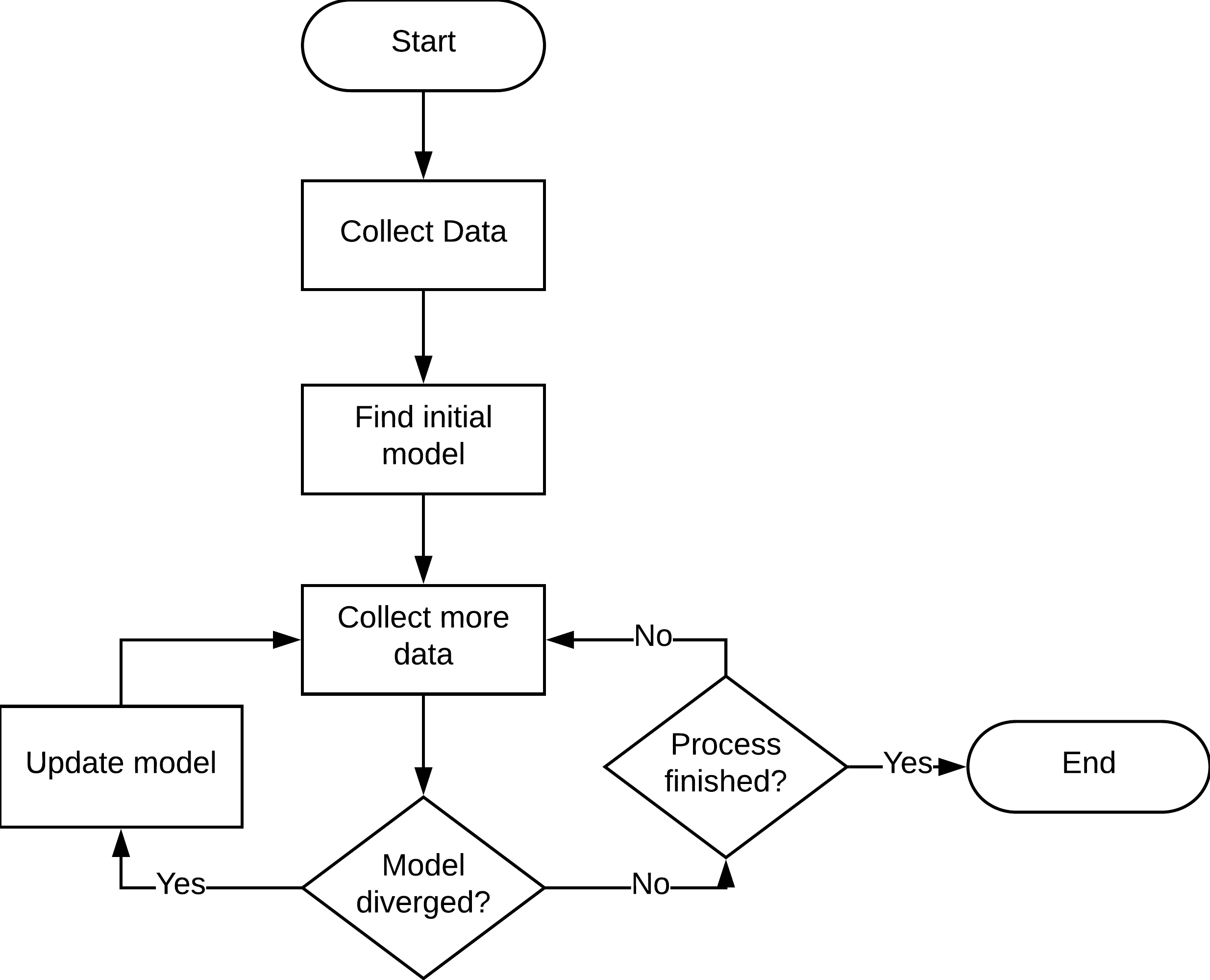}
    \centering
    \vspace{-.2in}
    \caption{Adaptive SINDy flow chart. For an initial model and hyper parameter selection, a gridsearch is conducted. Next, we apply a predictor corrector scheme checking every $t_{\text{error}}$ for model divergence using estimated Lyapunov time, and eventually update the model in a two step fashion.}
    \label{fig:flow-chart}
        \vspace{-.15in}
\end{figure}

Combinations of these changes, such as a simultaneous addition and deletion, are more challenging and will also be explored.  
This approach is known as \emph{abrupt-SINDy}, and it is depicted schematically in Fig.~\ref{fig:flow-chart}.

\subsection{Baseline model}
First, we must identify a baseline SINDy model, and we use a gridsearch to determine the optimal hyper parameter selection. 
In gridsearch, all combinations of hyper parameters are tested and the best performing set is selected. 
This search is only performed once, locking in hyper parameters for future updates. 
The baseline model is characterized by the sparse coefficients in $\boldsymbol{\Xi}_0$. 

\subsection{Detecting model divergence}
\label{sec:divergence}

It is essential to rapidly detect any change in the model, and we employ a classical predictor-corrector scheme~\cite{gershenfeld1999nature}. 
The predictor step is performed over a time $\tau_{\text{pred}}$ in the interval $t,t+\tau_{\text{pred}}$ using the model valid at time $t$. 
The divergence of the predicted and measured state is computed at $t+\tau$ as $\|\Delta {\bf x}\| = \| \hat{{\bf x}}(t+\tau) - {\bf x}(t+\tau) \|$, where $\hat{{\bf x}}$ is the prediction and ${\bf x}$ is the measurement. 
The idea is to identify when the model and the measurement diverge faster than predicted by the dynamics of the system.  
For a chaotic system, the divergence of a trajectory is measured by the largest Lyapunov exponent of the system~\cite{ott2002chaos}, although a wealth of similar measures have been suggested~\cite{scholl2008handbook}. The Lyapunov exponent is defined as
\begin{equation}
 \lambda = \lim_{\tau \to \infty} \lim_{\Delta {\bf x}(t_0) \to 0} \frac{\left\langle \log \left( \frac{ \Delta {\bf x}(t_0+\tau) }{ \Delta {\bf x}({t_0}) }\right) \right\rangle}{\tau},
\end{equation}
and its inverse sets the fastest time scale. Here, the analogy of ensemble and time average is used, more precisely the local, finite-time  equivalent \cite{lai2011transient,ding2007nonlinear}. 
An improvement can be achieved by exploiting an ensemble, e.g. by adding noise to the state ${\bf x}$ that corresponds to the given measurement accuracy. 
Since we know the dynamical system for the prediction step, the Lyapunov exponent is determined by evolving the tangent space with the system \cite{kantz1994robust,pikovsky1998dynamic}. 

In our detection algorithm, we fix a fluctuation tolerance $\Delta {\bf x}$ and measure if the divergence time we find deviates from the expectation. If data are noisy, this tolerance must be significantly larger than the the typical fluctuation scale of the noise.
Formally, the model and measurements diverge if the time-scale given by the local Lyapunov exponent and prediction horizon disagree. The local Lyapunov exponent is computed directly from the eigenvalues of the dynamical system~\cite{kantz1994robust,pikovsky1998dynamic}. The prediction horizon $T(t)$ is the first passage time where prediction $\hat{{\bf x}}(t + \Delta t)$ with initial condition ${\bf x}(t)$ and and measurement ${\bf x}(t + \Delta t)$ differ by more than $\Delta {\bf x}$:
\begin{equation}
    T(t) = \argmax_{\Delta t} \norm{\hat{{\bf x}}(t + \Delta t) - {\bf x}(t + \Delta t)} < \norm{\Delta {\bf x}}.
    \label{eq:horizon}
\end{equation}

\begin{figure}[t]
\vspace{-.1in}
    \includegraphics[width=\columnwidth]{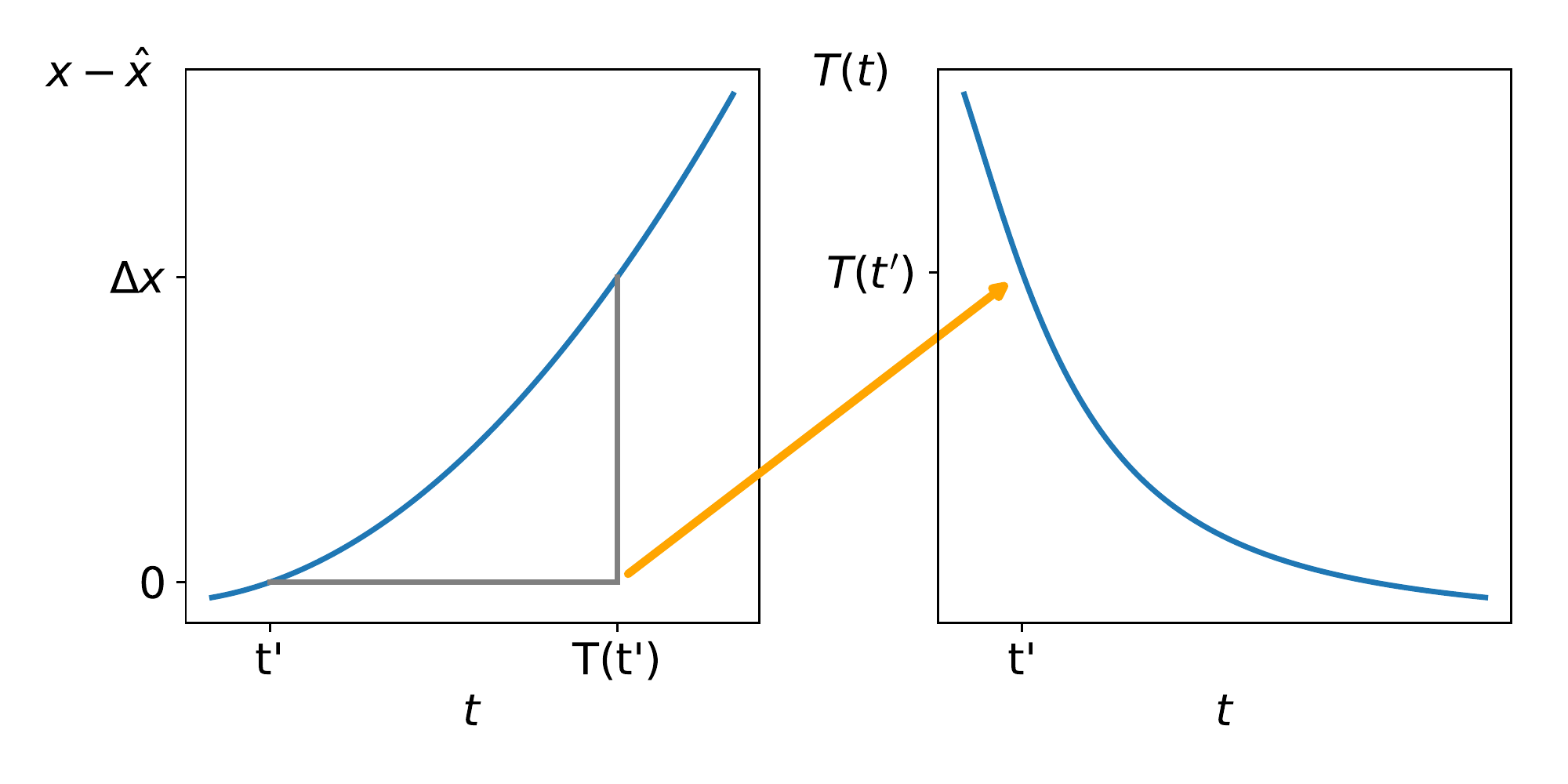}
    \centering
    \vspace{-.35in}
    \caption{Sketch of the prediction horizon estimation. We use the observation ${\bf x}(t)$ as initial condition for the current model. Integration gives $\hat{{\bf x}}(t)$. The prediction horizon $T(t)$ is calculated according to Eq.~\eqref{eq:horizon}. The prediction horizon is a function of time and the current model. It indicates divergence of model and observation. For details see text.}
    \label{fig:divergence}
\end{figure}

Analogous to the local Lyapunov exponent, we compute the ratio $\log \left\| \Delta {\bf x}(t_0+\tau)/\Delta {\bf x}(t_0) \right \|$ as a measure for the divergence based on the measurement.
For the model, we compute the local Lyapunov exponent as the average maximum eigenvalue $\bar{\lambda}(t) = \langle \lambda (t') \rangle_{t' \in [t, t+T]}$ with $\lambda (t) = \max(\lambda_i(t))$ and $\lambda_i v_i(t) = \left. {\partial f_j}/{\partial {\bf x}_k} \right| _{{\bf x}(t)} v_i(t)$.
Thus we compare the expected and observed trajectory divergence.
Model and measurement have diverged at time $t$ if the model time scale and the measured one differ:
\begin{equation}
    \bar{\lambda}(t) > \alpha \dfrac{\log(\Delta {\bf x}) - \log(\bar{\Delta}(t))}{T(t)}.
    \label{eq:estlyap}
\end{equation}
If the model is not chaotic, but the measurement is chaotic, one must invert the inequality, as in Fig.~\ref{fig:divergence}. The empirical factor $\alpha$ accounts for finite-time statistics.  

This method depends heavily on the particular system under investigation, including the dynamics, time scales, and sampling rate. 
In a practical implementation, these considerations must be handled carefully and automatically.   
It is important to note that we are able to formulate the divergence in terms of dynamical systems theory, because our model \textit{is} a dynamical system, in other cases, such as artificial neural networks, this is not possible due to the limited mathematical framework.

\subsection{Adaptive model fitting}
After a change is detected, the following procedure is implemented to rapidly recover the model:  
\begin{enumerate}
    \item First, the new data is regressed onto the existing sparse structure $\boldsymbol{\Xi}_0$ to identify varying parameters. 
    \item Next, we identify deletions of terms by performing the sparse regression on the sparse columns of $\boldsymbol{\Theta}$ that correspond to nonzero rows in $\boldsymbol{\Xi}_0$.  This is more efficient than identifying a new model, as we only seek to delete existing terms from the model.  
    \item Finally, if there is still a residual error, then a sparse model is fit for this error in the inactive columns of $\boldsymbol{\Theta}$ that correspond to zero rows in $\boldsymbol{\Xi}_0$.  In this way, new terms may be added to the model.  
\end{enumerate}
If the residual is sufficiently small after any step, the procedure ends.  
Alternatively, the procedure may be iterated until convergence.  
We are solving smaller regression problems by restricting our attention to subsets of the columns of $\boldsymbol{\Theta}$. 
These smaller regressions require less data and are less computationally expensive~\cite{Li2016}, compared to fitting a new model.  
The deletion-addition procedure is performed after a model divergence is detected, using new transient data collected in an interval of size $t_{\text{update}}$. 
\section{Results}\label{sec:gridsearch}
In this section, we describe the results of the abrupt-SINDy framework on dynamical systems with abrupt changes, including parameter variation, deletion of terms, and addition of terms.
The proposed algorithm is compared against the original SINDy algorithm, which is used to identify a new model from scratch, in terms of data required, computational time, and model accuracy.

In each case, we begin by running a gridsearch algorithm\cite{pedregosa2011scikit}\footnote{The user manual is located at \url{http://scikit-learn.org/stable/modules/generated/sklearn.model_selection.GridSearchCV.html}.} to identify the main parameters: $\alpha$, the ridge regression regularization parameter; $\gamma$, the thresholding parameter; $n_{\text{degree}}$, the maximum degree of the polynomial feature transformation; and $n_{\text{fold}}$, the number of cross-validation runs. For scoring we use the explained variance score and conduct a five-fold cross validation for each point in the $(\alpha, \gamma, n_{\text{degree}})$ parameter grid.

\begin{table}[]
\centering
\label{tab:parameters_gridsearch}
\begin{ruledtabular}

\begin{tabular}{l|l}
    Parameter & Value \\ \hline
    $\alpha$ & $0,0.2,0.4,0.6,0.8,0.95$ \\
    $\gamma $ & $0.1,0.2,0.4$\\
    $n_{\text{degree}}$ & $2,3$ \\
    $n_{\text{fold}}$ & $5$ \\
    Seed &  $42$ \\
    CV & k-fold \\
    Score & explained variance score
\end{tabular}

\end{ruledtabular}
\caption{Parameters for the grid search.}

\end{table}

\vspace{-.1in}
\subsection{Lorenz system}\label{Sec:Lorenz}
\vspace{-.1in}
The Lorenz system is a well-studied, and highly simplified, conceptual model for atmospheric circulation~\cite{Lorenz1963jas}:
\begin{equation}
 \begin{aligned}
     \dot{x} &= \sigma(y -x) \\
     \dot{y} &= \rho x - xz - y \\
     \dot{z} &= xy -\beta z
 \end{aligned}
\end{equation}
where the parameter $\rho$ represents the heating of the atmosphere, corresponding to the Rayleigh number, $\sigma$ corresponds to Prandtl number, and $\beta$ to the aspect ratio~\cite{strogatz2014nonlinear}.
The parameters are set to $\rho=28$, $\beta=8/3$, $\sigma=10$.

\begin{figure}[t]
    \centering
    \includegraphics[width=\columnwidth]{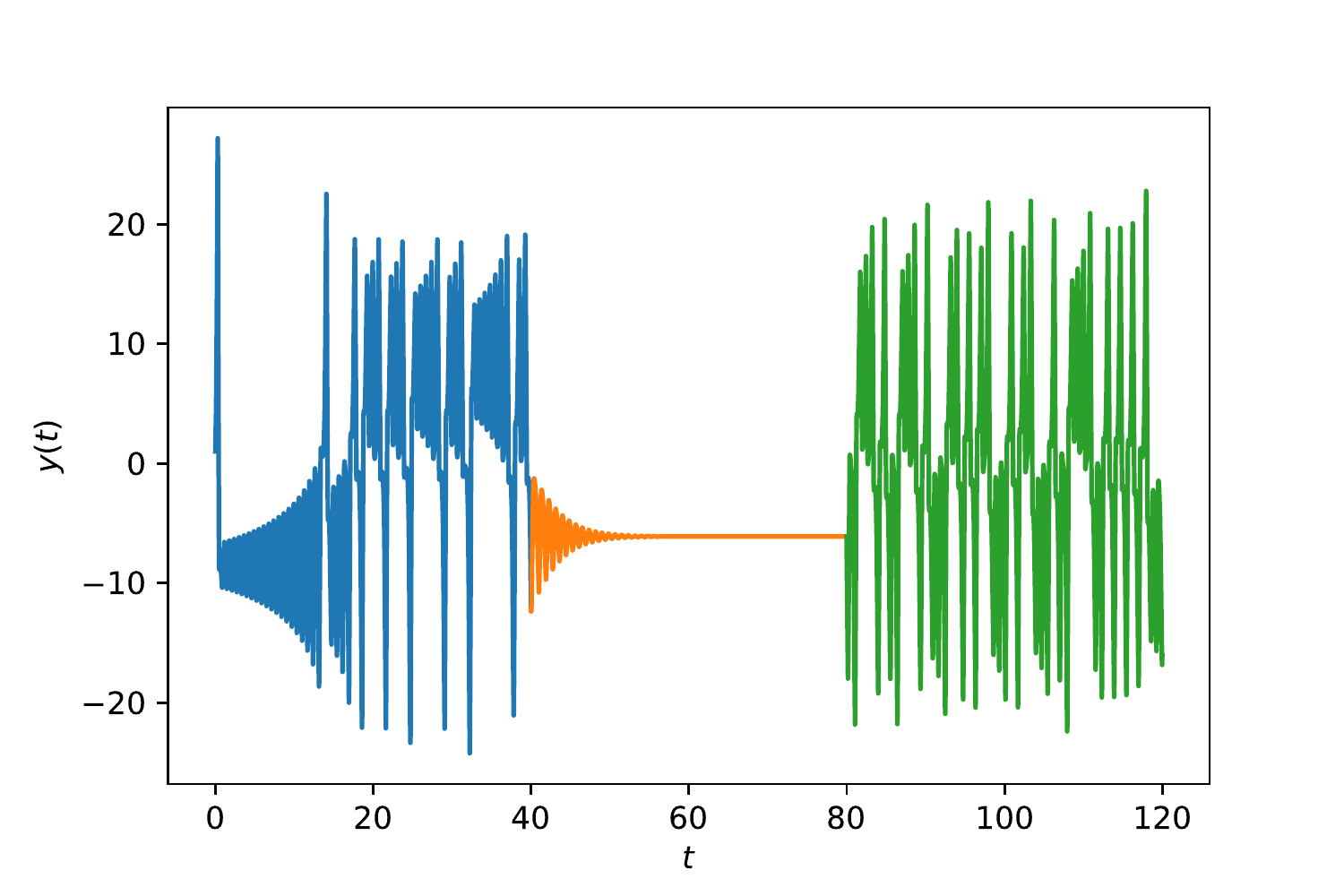}
    \vspace{-.325in}
    \caption{Time-series of the $y$ coordinate of the Lorenz system. The blue and green segments correspond to the system parameters $\sigma=10, \rho=28, \beta=\frac{8}{3}$. The orange segment from $t=40$ and $t=80$ corresponds to the modified parameter $\rho=15$. The initial condition is $\bx_0 = (1, 1, 1)$.}
    \label{fig:lorenz_y}
\end{figure}

\begin{figure}[t]
    \centering
    \includegraphics[width=\columnwidth]{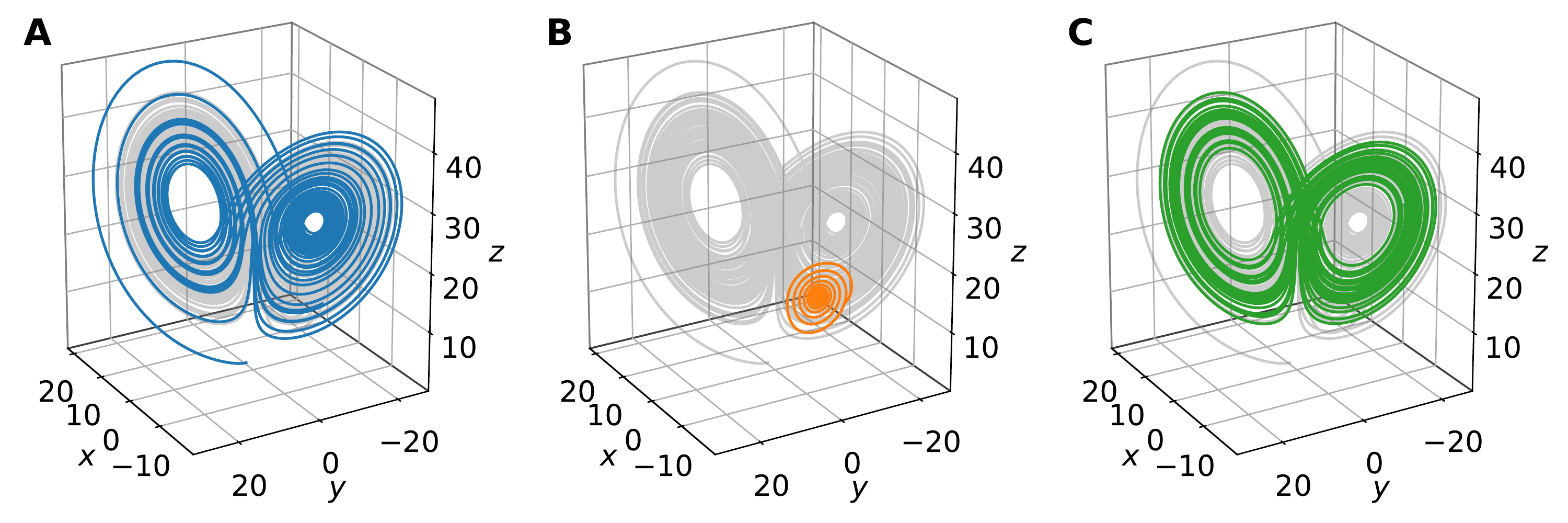}
    \vspace{-.3in}
    \caption{Lorenz system: Colors and parameters as in Fig.~\ref{fig:lorenz_y}. In \textbf{A}, \textbf{B}, and \textbf{C} we show the first, second, and third segments of the trajectory in color with the concatenated trajectory in grey. The system changes from a butterfly attractor to a stable fixed point and back to a butterfly attractor.}
    \label{fig:lorenz_3d}
        \vspace{-.1in}
\end{figure}

In the following we integrate the system numerically to produce a reference data set.
We deliberately change the parameter $\rho$ at $t=40$ to $\rho=15$ and at $t=80$ back to $\rho=28$, as shown in Figs.~\ref{fig:lorenz_y} and~\ref{fig:lorenz_3d}.
These parametric changes lead to a bifurcation in the dynamics, and they are detected quickly.
The subsequent adapted parameters are accurately detected up to two digits, as shown in Table~\ref{tab:lorenz}.
Because we are identifying the sparse model structure on a normalized library $\boldsymbol{\Theta}$, with zero mean and unit variance, we must de-bias the parameter estimates by computing a least-squares regression onto the sparse model structure in the original unnormalized variables.
Otherwise, computing the least-squares regression in the normalized library, as is typically recommended in machine learning, would result in non-physical constant terms in the original unnormalized coordinates.

\begin{figure}[t]
\centering
\vspace{-.1in}
    \includegraphics[width=\columnwidth]{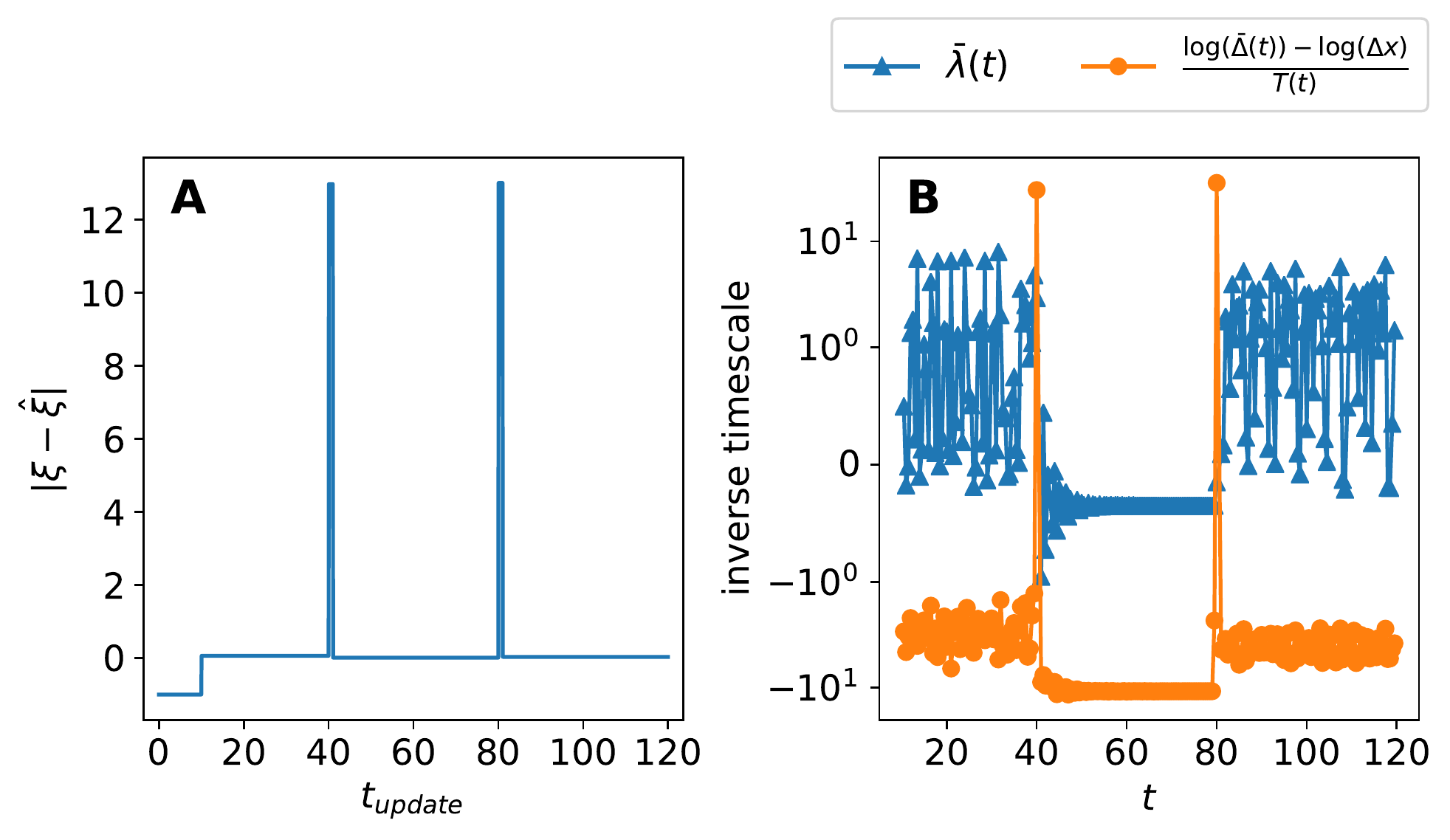}
    \centering
     \vspace{-.3in}
    \caption{Lorenz system: \textbf{A} We show the model accuracy over time. For coefficients, see Table ~\ref{tab:lorenz}. For $t\leq 10$ no model is available and $\|{\boldsymbol{\xi}} - \hat{{\boldsymbol{\xi}}}\| = -1$. At both switch points, $t=40$ and $t=80$, $t_{\text{update}} = 1$ is needed to update the model. During this interval, a fallback solution, e.g. DMD could be implemented. Note that the accuracy metric requires knowledge about the ground truth and thus is only available in a hindcast scenario. \textbf{B} Evaluation of Eq.~\eqref{eq:horizon}. At both switch points, $t=40$ and $t=80$, we quickly detect the divergence of model and measurement. The parameters are $t_{\text{model}}= 10, t_{\text{update}}= 1, t_{\text{error}}=0.5,$ and  $\Delta {\bf x}= 1.0$.}
    \label{fig:lorenz_horizon}
\end{figure}

Abrupt changes to the system parameters are detected using the prediction horizon from Eq.~\eqref{eq:horizon}.
When the system changes, the prediction horizon of the system should decrease, with smaller horizon corresponding to a more serious change.
Conversely, the inverse time, corresponding to the Lyapunov exponent, should diverge.
Figure~\ref{fig:lorenz_horizon} exhibits this expected behavior.
After a change is detected the model is rapidly recovered as shown in Table~\ref{tab:lorenz}.
It is important to confirm that the updated model accurately represents the structure of the true dynamics.
Figure~\ref{fig:lorenz_horizon} shows the norm of the model coefficients, $\|{\boldsymbol{\xi}} - \hat{{\boldsymbol{\xi}}}\|$, which is a measure of the distance between the estimated and true systems.
Except for a short time ($t_{\text{update}}=1$) after the abrupt change, the identified model closely agrees with the true model.

\begin{table}
    \centering
    \begin{ruledtabular}
    \begin{tabular}{rrl}
\toprule
 $t_{\text{detected}}$ &  $t_{\text{update}}$ &                                                                                                       Equations \\ \hline
\midrule
 0.00 &  10.0 &  $\begin{aligned}\dot{x} & = -10.0x+10.0y\\\dot{y} & = 27.96x-0.99y-1.0xz\\\dot{z} & = -2.67z+1.0xy\end{aligned}$ \\ \hline
 40.01 &  41.0 &  $\begin{aligned}\dot{x} & = -10.0x+10.0y\\\dot{y} & = 15.0x-1.0y-1.0xz\\\dot{z} & = -2.67z+1.0xy\end{aligned}$ \\ \hline
 80.02 &  81.0 &  $\begin{aligned}\dot{x} & = -10.0x+10.0y\\\dot{y} & = 27.98x-1.0y-1.0xz\\\dot{z} & = -2.67z+1.0xy\end{aligned}$ \\
\bottomrule
\end{tabular}

    \end{ruledtabular}
    \vspace{-.1in}
    \caption{Lorenz system: detection and update times, along with identified equations. The detection time coincides up to the second digit with the true switching time. The rapidly identified model agrees well with the true model structure and parameters. Coefficients are rounded to the second digit.}
    \vspace{-.15in}
    \label{tab:lorenz}
\end{table}

\vspace{-.15in}
\subsubsection{Effects of noise and data volume}
\vspace{-.1in}
An important set of practical considerations include how noise and the amount of data influence the speed of change detection and the accuracy of subsequent model recovery.
Both the noise robustness and the amount of data required will change for a new problem, and here we report trends for this specific case.
In addition, the amount of data required is also related to the sampling rate, which is the subject of ongoing investigation; in some cases, higher sampling time may even degrade model performance due to numerical effects~\cite{Ahnert-Abel-2007}.

\begin{figure}[t]
\vspace{-.15in}
    \includegraphics[width=\columnwidth]{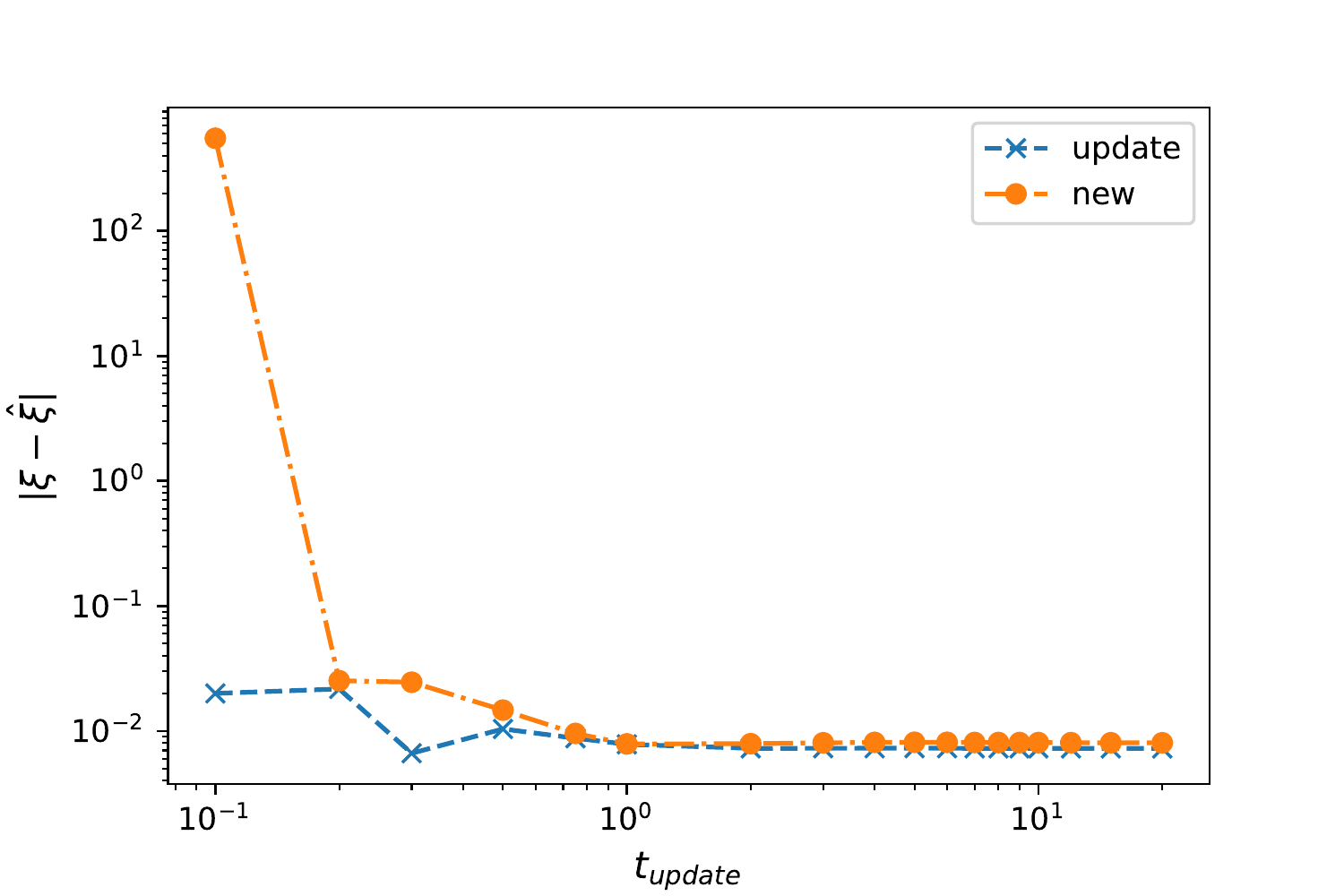}
    \centering
        \vspace{-.25in}
    \caption{Lorenz system: We show the model accuracy versus the amount of data used to update (blue $\times$) or re-fit (orange dot) respectively. Data is collected from in the interval $[40, 40 + t_{\text{update}}]$ just after the first change of the system dynamics. The number of data points for $t_{\text{update}} =0.1$ are $N=25$, for $t_{\text{update}} =10$ we have 2500 points. At $t_{\text{update}} \simeq 1$, updating and re-fitting methods become comparable. However, for smaller update times, or less data, respectively, the fraction of transient data becomes too small for identifying the exact model from scratch. Updating the model needs less data for the same accuracy or achieves higher accuracy with the same amount of data.}
    \label{fig:lorenz_update_scaling}
\end{figure}

Figure~\ref{fig:lorenz_update_scaling} shows the model fit following the abrupt change, comparing both the abrupt-SINDy method, which uses information about the existing model structure, and the standard SINDy method, which re-identifies the model from scratch following a detected change.
In this figure, the model quality is shown as a function of the amount of data collected after the change.

The abrupt-SINDy model is able to identify more accurate models in a very short amount of time, given by $t_{\text{update}}\approx 0.1$.
At this point, the standard SINDy method shows comparable error, however for even smaller times, the data are no longer sufficient for the conventional method.
Since the adaptive method starts near the optimal solution, larger data sets do not degrade the model, which was an unexpected additional advantage.

\begin{figure}[t]
    \includegraphics[width=\columnwidth]{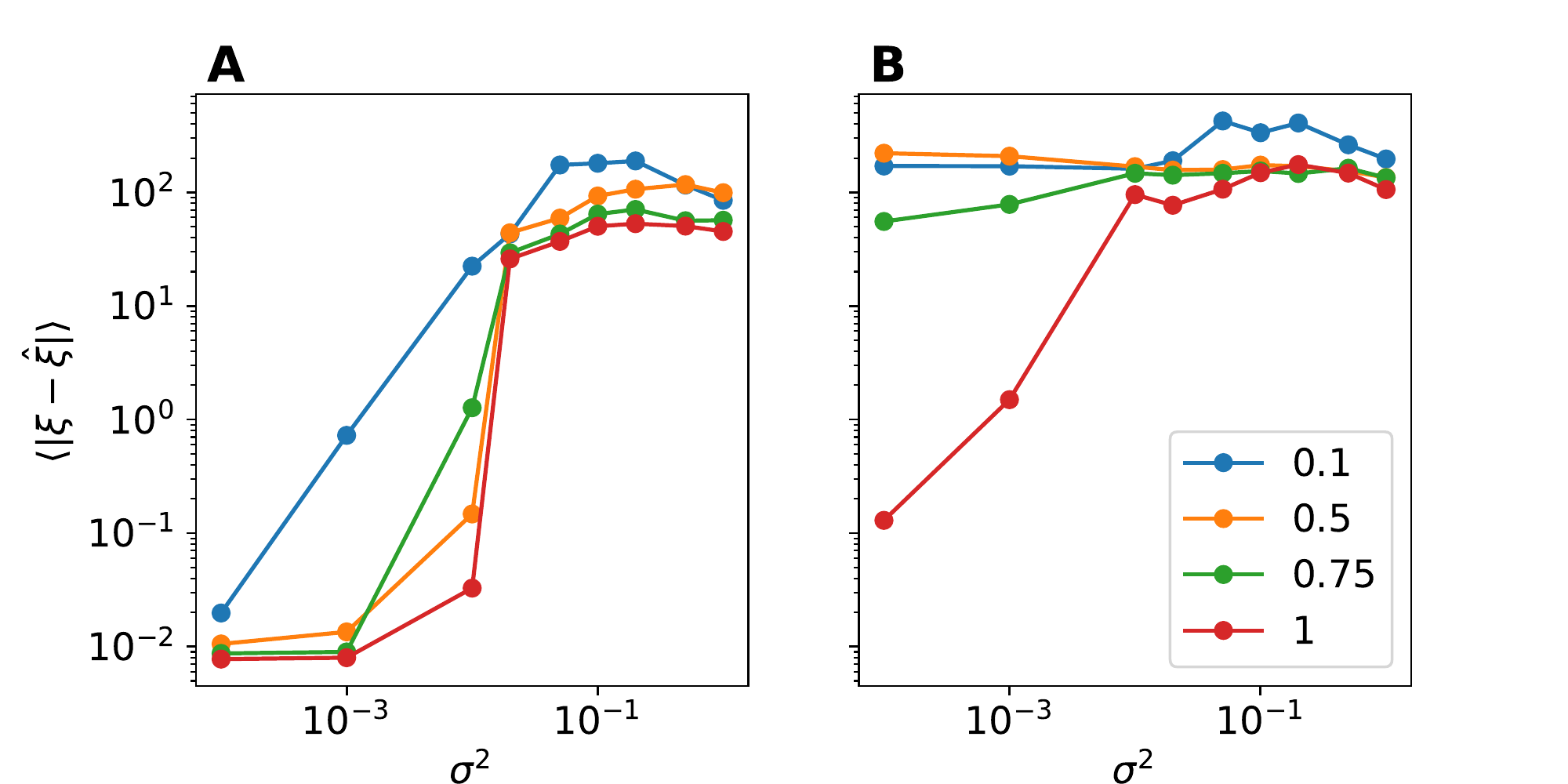}
    \centering
        \vspace{-.25in}
    \caption{Lorenz system: We show the noise robustness of model accuracy. In A) we use the previous knowledge and update the model; in B) we make a new fit only re-using the previously discovered hyper-parameters. The curves are parametrized by $t_{\text{update}}$, c.f. Fig.~\ref{fig:lorenz_update_scaling}. The accuracy measure is very noise sensitive, as distinction between library functions gets lost. At a signal to noise ration of approximately $1$, no accurate model can be obtained with either model. At lower noise ratios, updating the model achieves higher accuracy (the library is smaller). In both cases, accuracy scales approximately logarithmically with $t_{\text{update}}$.}
    \label{fig:lorenz_noise}
\end{figure}

Figure~\ref{fig:lorenz_noise} explores the effect of additive noise on the derivative on the abrupt-SINDy and standard SINDy algorithms.
Note that in practice noise will typically be added to the measurement of $\bx$, as in the original SINDy algorithm~\cite{Brunton2016pnas}, requiring a denoising derivative~\cite{Ahnert-Abel-2007,Chartrand2011isrnam}; however, simple additive noise on the derivative is useful to investigate the robustness of the regression procedure.
Abrupt-SINDy has considerably higher noise tolerance than the standard algorithm, as it must identify fewer unknown coefficients.
In fact, it is able to handle approximately an order of magnitude more noise before failing to identify a model.
Generally, increasing the volume of data collection improves the model. 
The critical point in the abrupt-SINDy curves corresponds to when small but dynamically important terms are mis-identified as a result of insufficient signal-to-noise.
Although the noise and chaotic signal cannot be easily distinguished for small signal-to-noise, it may be possible to distinguish between them using a spectral analysis, since chaos yields red noise in contrast to the white additive noise.

\subsection{Van der Pol}

\begin{figure}[t]
\centering
\vspace{-.2in}
    \includegraphics[width=\columnwidth]{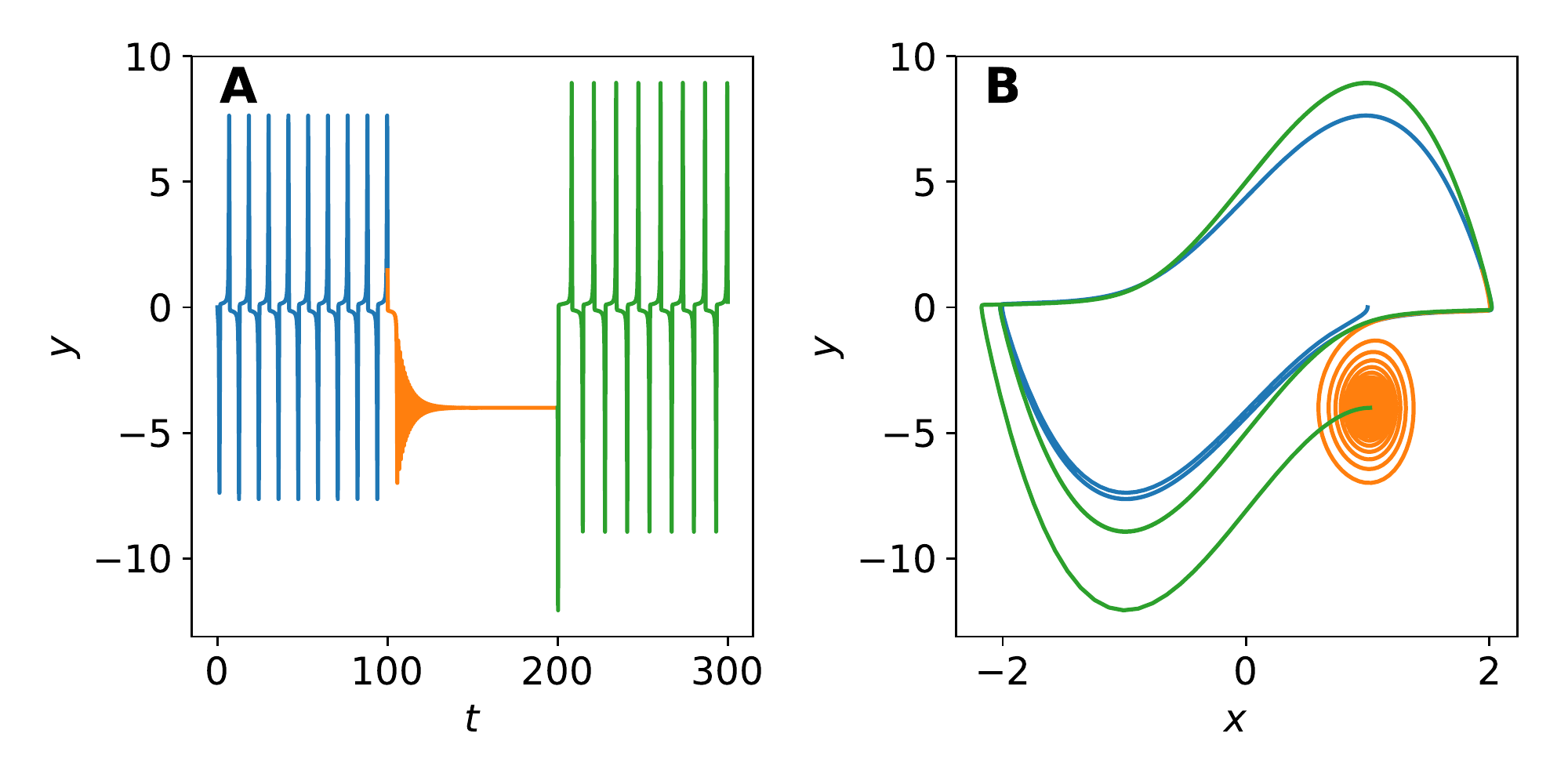}
        \vspace{-.3in}
    \caption{Van der Pol system with parameters $\mu=5, \alpha=0$ (blue), $\mu=7.5, \alpha=-0.25$ (orange), $\mu=6.0, \alpha=0$ (green). \textbf{A}: time evolution of the $y$-coordinate. \textbf{B} phase-space-trajectory $x, y$.}
    \label{fig:vdp}
\end{figure}

As a second example, we consider the famous nonlinear Van der Pol oscillator~\cite{van1920theory}.
We include additional quadratic nonlinearities $\alpha x^2$ and $\alpha y^2$ to study the ability of our method to capture structural changes when these terms are added and removed abruptly.
This example focuses on the important class of periodic phenomena, in contrast to the chaotic Lorenz dynamics.
The modified Van der Pol oscillator is described by the following equations:
\begin{equation}
    \begin{aligned}
     \dot{x} &= y - \alpha y^2 \\
     \dot{y} &= \mu (1 - x^2)y - x + \alpha x^2\;,\\
 \end{aligned}
\end{equation}
where $\mu > 0$ is a parameter controlling the nonlinear damping, and $\alpha$ parameterizes the additional quadratic nonlinearity.
The reference data set is shown in Fig.~\ref{fig:vdp}, with $\mu=7.5$ and $\alpha=0$ for $t \in [0, 100]$, which results in a canonical periodic orbit.
At $t=100$ we introduce a structural change, switching on the quadratic nonlinearity ($\alpha = -0.25$), and driving the system to a stable fixed point.
We also modify the parameter $\mu$, setting it to $\mu=6.0$.
Finally, at $t=200$, we switch off the additional nonlinearity ($\alpha = 0$) and keep $\mu = 6$.


\begin{figure}[t]
\vspace{-.1in}
    \includegraphics[width=\columnwidth]{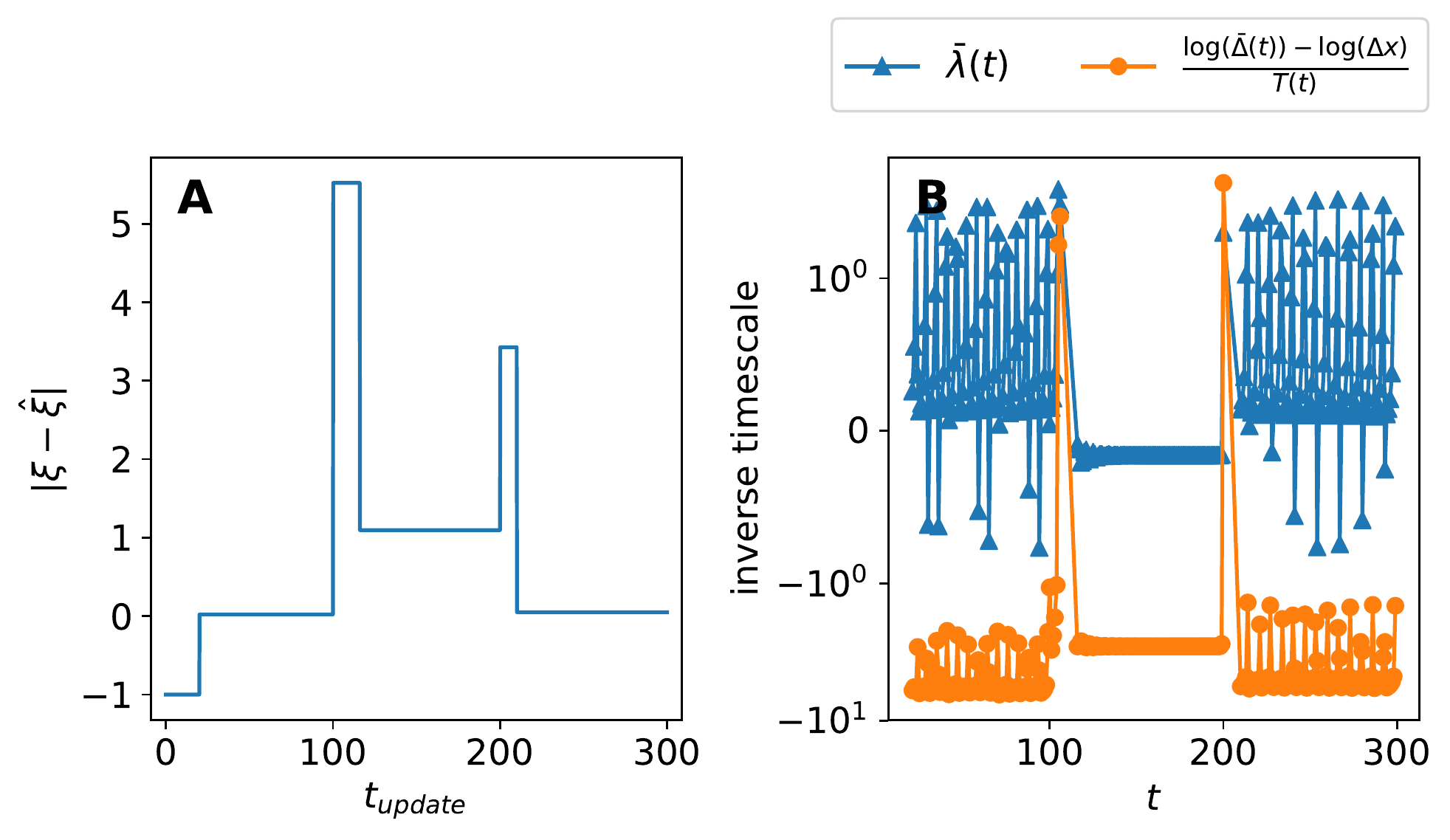}
    \centering
    \vspace{-.275in}
    \caption{Van der Pol system: Evaluation of Eq.~\eqref{eq:horizon}. Parameters: $t_{\text{model}}= 20, t_{\text{update}}=10, t_{\text{error}}=1, \Delta x= 1.5$.}
    \label{fig:vdp_error_horizon}
\end{figure}

Table~\ref{tab:vdp} shows the corresponding models recovered using the abrupt-SINDy method.
The change is detected using the Lyapunov time defined in Eq.~\eqref{eq:estlyap}, as shown in Fig.~\ref{fig:vdp_error_horizon}.
Again, the estimated Lyapunov time (Fig.~\ref{fig:vdp_error_horizon} \textbf{B}) captures the the changes in the model, which correspond to peaks in structural model error (Fig.~\ref{fig:vdp_error_horizon} \textbf{A}).
While the first and third stage are indeed identified correctly, the term $-1.25 x$ is preferred over $-x - 0.25x^2$ in the sparse estimate for $\dot{y}$ in the orange trajectory.
However, since both terms look similar near the fixed point at $x \sim 1$, this describes the dynamics well.
This type of mis-identification often occurs in data mining when features are highly correlated~\cite{Li2016} and is more related to sparse regression in general than the proposed abrupt-SINDy.
For dynamic system identification, the correct nonlinearity could be resolved by obtaining more transient data, i.e. by perturbing the system through actuation.
However, this model may be sufficient for control while a more accurate model is identified.

\begin{table}
    \centering
    \begin{ruledtabular}
    \begin{tabular}{rrl}
\toprule
 $t_{\text{detected}}$ &  $t_{\text{update}}$ &                                                                                  Equations \\ \hline
\midrule
 0.00 &  20.01 &  $\begin{aligned}\dot{x} & = 1.0y\\\dot{y} & = -1.0x+4.99y-4.99x^2y\end{aligned}$ \\ \hline
 106.39 &  116.00 &  $\begin{aligned}\dot{x} & = 0.99y+0.25y^2\\\dot{y} & = -1.26x+7.46y-7.46x^2y\end{aligned}$ \\ \hline
 200.12 &  210.00 &  $\begin{aligned}\dot{x} & = 1.0y\\\dot{y} & = -1.0x+5.98y-5.98x^2y\end{aligned}$ \\
\bottomrule
\end{tabular}

    \end{ruledtabular}
    \vspace{-.1in}
    \caption{Van der Pol system: Summary of the discovered equations. Coefficients are rounded to the second digit.}
    \label{tab:vdp}
    \vspace{-.1in}
\end{table}

\section{Conclusions}
In this work, we develop an adaptive nonlinear system identification strategy designed to rapidly recover nonlinear models from limited data following an abrupt change to the system dynamics.
The sparse identification of nonlinear dynamics (SINDy) framework is ideal for change detection and model recovery, as it relies on parsimony to select the fewest active terms required to model the dynamics.
In our adaptive abrupt-SINDy method, we rely on previously identified models to identify the fewest \emph{changes} required to recover the model.
This modified algorithm is shown to be highly effective at model recovery following an abrupt change, requiring less data, less computation time, and having improved noise robustness over identifying a new model from scratch.
The abrupt-SINDy method is demonstrated on several numerical examples exhibiting chaotic dynamics and periodic dynamics, as well as parametric and structural changes, enabling real-time model recovery.


There are limitations of the method which can be addressed by several promising directions that may be pursued to improve the abrupt-SINDy method:
\begin{enumerate}
 \item \textbf{Fallback models:} In the current implementation, after a change has been detected, the old model will be used until enough data is collected to identify a new model.
The dynamic mode decomposition~\cite{Kutz2016book} provides an alternative fallback model, that may be identified rapidly with even less data.
Additionally, instead of relying on a sparse update to the current model, it is sensible to also maintain a library of past models for rapid characterization~\cite{Brunton2014siads}.
 \item \textbf{Hyperparameterization:} In the initial prototype, the hyper-parameters $\Delta_x$ and $t_{\text{update}}$ are fixed.
 Over time, an improved algorithm may learn and adapt optimal hyper-parameters.
\item \textbf{Comprehensive Lyapunov time estimation:} According to Eq.\eqref{eq:horizon}, the Lyapunov time $T(t | \Delta {\bf x})$ is estimated for a fixed $\Delta {\bf x}$.
Estimating the time for a range of values, i.e. $\Delta {\bf x} \in (0, \Delta {\bf x}_{\max}]$, will be more robust and may provide a richer analysis without requiring additional data. Further investigation must be made into the case of chaotic systems, where the numerical calculation of the Lyapunov exponent may fail to reveal divergence due to the fact of simple averaging over time. 
Because of the importance of the detection of model divergence, this is a particularly important area of future research.  
\item \textbf{Advanced optimization and objectives:} Looking forward, advanced optimization techniques may be used to further improve the adaptation to system changes.
Depending on the system, other objectives may be optimized, either by including regularization or in a multi-objective optimization.
\end{enumerate}

The proposed abrupt-SINDy framework is promising for the real-time recovery of nonlinear models following abrupt changes. 
It will be interesting to compare with other recent algorithms that learn local dynamics for control in response to abrupt changes~\cite{Ornik2017arxiv}.   
Future work will be required to demonstrate this method on more sophisticated engineering problems and to incorporate it in controllers.  
To understand the limitations for practical use, many further studies are needed, it will be particularly useful to test this method on a real experiment.
The abrupt-SINDy modeling framework may also help inform current rapid learning strategies in neural network architectures~\cite{fei2006one,vinyals2016matching,delahunt2018putting}, potentially allowing dynamical systems methods to inform rapid training paradigms in deep learning.


\section*{Acknowledgements}
MQ was supported by a fellowship within the FITweltweit program of the German Academic Exchange Service (DAAD).
MQ and MA acknowledge support by the European Erasmus SME/HPC project (588372-EPP-1-2017-1-IE-EPPKA2-KA).
SLB acknowledges support from the ARO and AFOSR Young Investigator Programs (ARO W911NF-17-1-0422 and AFOSR FA9550-18-1-0200).
SLB and JNK acknowledge support from DARPA (HR0011-16-C-0016).
We would like to thank the anonymous reviewers for their comments which helped to improve this manuscript.
We also acknowledge valuable discussions related to abrupt model recovery and programming wisdom with Dennis Bernstein, Karthik Duraisamy, Thomas Isele, Eurika Kaiser, and Hod Lipson.

\section*{References}
\bibliography{main}

\end{document}